\documentclass[a4paper]{jpconf}
\usepackage{amsmath,amssymb,amsfonts,amsthm,wasysym}
\usepackage{epsfig,graphics,graphicx,epstopdf}
\usepackage{array,booktabs,colortbl,colordvi,multirow}

\newcommand{\hc}{\mathrm{h.c.}}

\newcommand{\U}[1]{U(#1)}
\newcommand{\SU}[1]{SU(#1)}

\newcommand{\beq}{\begin{equation}}
\newcommand{\eeq}{\end{equation}}
\newcommand{\bal}{\begin{align}}
\newcommand{\eal}{\end{align}}
\newcommand{\refeq}[1]{\mbox{Eq.~(\ref{#1})}}
\newcommand{\nn}{\nonumber \\} 

\long\def\symbolfootnote[#1]#2{\begingroup%
\def\thefootnote{\fnsymbol{footnote}}\footnote[#1]{#2}\endgroup} 

\begin{document}
\title{Top couplings and top partners\footnote{Talk given by M.\ P\'erez-Victoria at Top 2012, Winchester, UK, September 16-21, 2012.}}

\author{J A Aguilar Saavedra and M P\'erez-Victoria}

\address{CAFPE and Departamento de F\'{\i}sica Te\'orica y del Cosmos\\Universidad de Granada, E-18071, Spain}

\ead{jaas@ugr.es,mpv@ugr.es}

\begin{abstract}
We review the model-independent description of the couplings of the top quark to the Higgs and gauge bosons in theories beyond the Standard Model. Then we examine these couplings in the case of arbitrary heavy vector-like quarks mixing with the third family. We also discuss the couplings of these top partners, and comment on implications for LHC searches. 
\end{abstract}

\section{Introduction}
The experiments at Tevatron and the LHC have allowed to determine with remarkable precission many properties of a resonance with mass around 173~GeV and width about 2~GeV, discovered at Tevatron in 1995. The properties measured so far are consistent with the ones of the particle called top quark: a spin-1/2 fermion with electric charge $Q_t=2/3$ that is a colour triplet and whose left-handed component is the weak isospin partner of the left-handed $b$ quark. A particle with these quantum numbers is actually required for anomaly cancellation, and furthermore the measured mass is compatible with electroweak precision tests. Therefore, it is safe to assume that the top-like resonance observed in hadron colliders is actually the top quark, as defined two sentences above. This by no means implies that the couplings of this particle are equal to the ones in the Standard Model (SM).\footnote{Still, we call it "top quark". Analogous experimental, theoretical and linguistic considerations may be applied to the recently discovered Higgs-like resonance, which will be called "Higgs boson" hereafter.} Indeed, the simple form of the SM top couplings follows not only from the SM gauge invariance and top quantum numbers, but also from renormalizability and the complete SM field content. In this talk we will examine top couplings in the absence of these last two hypotheses. Especifically, we first consider corrections to the top couplings from non-renormalizable operators constructed with the SM fields, and then study a renormalizable extension of the SM with heavy vector-like quarks. We will keep the requirement of invariance under the (spontaneously broken) SM gauge group, since the consequences of breaking this symmetry are rather disastrous for our understanding of nature. To keep the discussion as simple as possible, we will restrict ourselves to new physics that couples only to the third family in the interaction basis, and assume that electroweak breaking is induced by a light Higgs doublet, just as in the SM. Moreover, we focus on the trilinear top interactions.

\section{General top couplings}
Let us study the trilinear couplings of the top quark in the presence of arbitrary heavy new physics at a scale $\Lambda$, which we assume to be of the decoupling kind. The effective Lagrangian describing processes at energies smaller than the scale $\Lambda$ can be expanded in a power series
\beq
\mathcal{L}^{\mathrm{eff}}= \mathcal{L}^{(4)}+\frac{1}{\Lambda} \mathcal{L}^{(5)}+ \frac{1}{\Lambda^2} \mathcal{L}^{(6)}+\ldots ,
\eeq
where each $\mathcal{L}^{(n)}$ is a finite sum of local operators $\mathcal{O}_i^{(n)}$ of scaling dimension $n$, made of SM fields. These operators are required to be invariant under the full $\SU{3}_C\times \SU{2}_L \times \U{1}_Y$ gauge group. Note that if we imposed only color and electromagnetic symmetries, we would introduce unnecessary parameters and miss some model-independent relations between the couplings. The lowest-order term, $\mathcal{L}^{(4)}$, is just the SM Lagrangian. The first corrections to top couplings arise from $\mathcal{L}^{(6)}$ and change substantially the SM pattern: flavour-changing neutral currents (FCNC) and right-handed charged currents are allowed at the tree-level, the left-handed CKM matrix needs not be unitary and extra CP phases can appear~\cite{dAPVS1}. With our assumption of third-family-philic new physics, the mixing with the light families of quarks is suppressed by small CKM entries and can be neglected. $\mathcal{L}^{(6)}$ also contains four-fermion operators, which give important contributions to top pair production---see~\cite{Degrande,Delaunay,APV1,APV2} and the talk by J.\ Kamenik in this conference---but are not studied here. A convenient minimal set of operators contributing to trilinear interactions has been given in Refs.~\cite{JAminimalgauge,JAminimalHiggs}. Using these operators, we can write the most general effective interactions of the top quark with the gauge bosons and the Higgs, to order $\Lambda^{-2}$, in the following fashion:
\bal
& \mathcal{L}_{gtt}= -g_s \bar{t}\frac{\lambda^a}{2}\gamma^\mu t G^a_\mu - g_s \bar{t} \lambda^a \frac{i\sigma^{\mu\nu} q_\nu}{m_t} \left( d_V^g + i d_A^g \gamma_5\right) t G^a_\mu. \label{anomalous1st} \\
& \mathcal{L}_{\gamma tt}= -e Q_t \bar{t}\gamma^\mu t A_\mu - e \bar{t} \frac{i\sigma^{\mu\nu} q_\nu}{m_t} \left( d_V^{t\gamma} + i d_A^{t\gamma} \gamma_5\right) t A_\mu.\\
& \mathcal{L}_{W tb}= -\frac{g}{\sqrt{2}} \bar{b}\gamma^\mu \left(V_L P_L+V_R P_R \right) t W^-_\mu - \frac{g}{\sqrt{2}} \bar{b} \frac{i\sigma^{\mu\nu} q_\nu}{M_W} \left(g_L P_L + g_R P_R \right) t W^-_\mu + \hc \, . \\
& \mathcal{L}_{Z tt}= -\frac{g}{2c_W} \bar{t}\gamma^\mu \left(X^t_L P_L+X^t_R P_R -2 s_W^2 Q_t \right) t Z_\mu - \frac{g}{2c_W} \bar{t} \frac{i\sigma^{\mu\nu} q_\nu}{M_Z} \left(d_V^{tZ} + i d_A^{tZ} \gamma_5 \right) t Z_\mu . \\
& \mathcal{L}_{H tt}= -\frac{1}{\sqrt{2}} \bar{t} \left(Y_V+i Y_A \gamma_5 \right) t H . \\
& \mathcal{L}_{\gamma bb}= e Q_b \bar{b}\gamma^\mu b A_\mu + e \bar{b} \frac{i\sigma^{\mu\nu} q_\nu}{m_b} \left( d_V^{b\gamma} + i d_A^{b\gamma} \gamma_5\right) b A_\mu.\\
& \mathcal{L}_{Z bb}= \frac{g}{2c_W} \bar{b}\gamma^\mu \left(X^b_L P_L+X^b_R P_R +2 s_W^2 Q_b \right) b Z_\mu + \frac{g}{2c_W} \bar{b} \frac{i\sigma^{\mu\nu} q_\nu}{M_Z} \left(d_V^{bZ} + i d_A^{bZ} \gamma_5 \right) b Z_\mu . \label{anomalouslast}
\end{align}
We emphasize that this parametrization is completely general and valid not only on-shell, but also when the particles are off-shell or inside loops.\footnote{The effect in the amplitudes of other possible trilinear terms can be reproduced by the ones we have written plus four-fermion interactions, see \cite{JAminimalgauge} for more details.} We have also written the trilinear interactions of the b quark with the $Z$ boson and the photon because, as we will see shortly, they are connected to the top interactions.
At order $\Lambda^0$, the effective couplings above are as in the SM: $g_{L,R}$, $V_{R}$, $X^{t,b}_R$, $d_{V,A}^{g,t\gamma,tZ,b\gamma,bZ}$ and $Y_A$ vanish, while $X^t_L=X_L^b=1$, $V_L=V_{tb}^{\mathrm{SM}}\simeq 1$ and $Y_V=\sqrt{2}m_t/v$, with $v\simeq 246\,\mathrm{GeV}$ the Higgs vaccuum expectation value. At order $\Lambda^{-2}$, all the effective couplings, which we represent generically by $\kappa_j$, receive corrections of the form
\beq
\delta \kappa_j = \kappa_j-\kappa_j^\mathrm{SM} = A_j \frac{v^2}{\Lambda^2}, 
\eeq
where $A_j$ are dimensionless linear combinations of the operator coefficients. These corrections are called anomalous couplings. The gauge invariance of the operators implies the following relations:
\bal
& \delta X_L^t + \delta X_L^b = 2 \delta V_L , \label{tbrelation} \\
& c_W d^{tZ}+e s_W \frac{v}{m_t} d^{t\gamma} = g_R, \\
& c_W d^{bZ}+e s_W \frac{v}{m_b} d^{b\gamma} = g_L^*, 
\end{align}
where we have defined $d^{t,b\,Z,\gamma}=d_V^{t,b\,Z,\gamma} + i d_A^{t,b\,Z,\gamma}$.

Let us make two simple observations with important consequences in the analysis of top data. First, even though the anomalous coupling $\delta V_L$, which accounts for the loss of unitarity of a CKM matrix beyond the SM, can be taken to be real \cite{JAminimalHiggs}, its sign is not determined {\em a priori}. Actually, we give below examples of SM extensions giving rise to either sign. Therefore, $V_L$ can be smaller than, equal to or larger than 1. Requiring $V_L\leq 1$ (or $V_L\geq 1$) is a prior without any model-independent theoretical justification.

The second observation comes from relation \refeq{tbrelation}. Electroweak precision data ($R_b$ at the Z pole, in particular) put a tight limit on the anomalous coupling $\delta X_L^b$. Using this piece of information, we can approximate $\delta X_L^b\simeq 0$ \footnote{This condition occurs naturally in some models, see the examples below.} and conclude that
\beq
\delta X_L^t \simeq 2 \delta V_L.
\eeq
This means that, to a good approximation, the zero-momentum couplings of the top quark to the $W$ and $Z$ bosons depend on just one real parameter. In this way, the limit on $V_L$ from single top production can be translated into a limit on the coupling of the left-handed top to the $Z$ boson. Using the results in Ref.~\cite{AtlasVL}, we find $X_L^t\in [0.7,1.8]$.

To finish this section, we discuss the relevance of the different anomalous couplings and  their expected size. Because the $\delta \kappa_j$ are of order $\Lambda^{-2}$, their contribution to observables will be $O(\Lambda^{-2})$ when the amplitudes these insertions interfere with the SM ones, and $O(\Lambda^{-4})$ otherwise. For this reason, some authors drop non-interfering anomalous couplings from their analyses. However, it should be noted that these couplings contribute to observables that vanish to $O(\Lambda^{-2})$. Hence, they give the leading contribution to some genuinely new effects with small backgrounds. This, in many cases, compensates for the extra suppression. One example is given by the top helicity fraction $F_+$, which is very small in the SM. The leading corrections to $F_+$ arise from the anomalous couplings $V_R$ and $g_L$, and are $O(\Lambda^{-4})$ (plus $O(m_b^2\Lambda^{-2})$) \cite{JABernabeu}. Dimension-8 operators only contribute to such observables at order $\Lambda^{-8}$ and can be safely neglected. An additional reason for keeping all the anomalous couplings is that some extensions of the SM only produce the non-interfering type.

On the other hand, it is easy to show that none of the vertices above with a $\sigma^{\mu\nu}$ can be generated by any new physics at the classical level. Therefore, the corresponding couplings $d$ are expected to have a $1/(16\pi^2)$ suppression relative to the other anomalous couplings, as long as i) they arise from new physics at the same scale and ii) the couplings of the new particles to the SM ones are perturbative. These conditions are met, for instance, in the SM extensions considered in the next section. But once again, to keep full generality in a model-independent analysis all the anomalous couplings above should be included.

\section{Top partners}

The anomalous top couplings in the previous section can be generated at the classical level in theories that contain extra bosons mixing with the SM Higgs or gauge bosons and/or extra quarks that mix with the top or bottom quark. We study here extensions with extra quarks that couple to the third family. New chiral quarks, such as the ones in a 4th generation of fermions, are all but excluded by the new Higgs data~\cite{4gen1,4gen2} (besides having severe difficulties with electroweak precision observables and perturbativity). Therefore, we concentrate on the case of extra vector-like quarks, whose left-handed and right-handed components have the same gauge quantum numbers. These heavy particles appear at the TeV scale in several well motivated extensions of the SM, such as extra dimensions, composite Higgs and little Higgs theories. In order to modify the top couplings at the tree level, the new quarks must mix with the top or bottom quarks. This is only possible for the following multiplets \cite{dAPVS2}: 
\beq
\begin{array}{llll}
\mathbf{1}_{\frac{2}{3}} = T, & \mathbf{1}_{-\frac{1}{3}} = B, & \mathbf{2}_{\frac{1}{6}} = \left( \begin{array}{c} T \\ B \end{array} \right), 
&  \mathbf{2}_{\frac{7}{6}} = \left( \begin{array}{c} X \\ T \end{array} \right), \\
\mathbf{2}_{-\frac{5}{6}} = \left( \begin{array}{c} B \\ Y \end{array} \right),
& \mathbf{3}_{\frac{2}{3}} = \left( \begin{array}{c} X \\ T \\ B \end{array} \right),
& \mathbf{3}_{-\frac{1}{3}} = \left( \begin{array}{c} T \\ B \\ Y \end{array} \right) .
\end{array}
\eeq
The electric charge is 2/3, -1/3, 5/3 and -4/3 for the components denoted by $T$, $B$, $X$ and $Y$, respectively. These vector-like fermions can have a gauge invariant Dirac mass $M$, and decouple in the limit $M\to \infty$. They can couple to the SM quarks via Yukawa interactions
of the form
\bal
& -\lambda_Q \bar{Q}_{R} \phi q_{L} + \hc, ~~ \mbox{if $Q$ is a singlet or triplet}, \nn
& -\lambda_Q \bar{Q}_L \phi q_R + \hc, ~~ \mbox{if $Q$ is a doublet} \label{Yukawa}.
\end{align}
$Q_{L,R}$ are the two chiral components of the vector-like multiplet $Q$, $q_{L,R}$ denote, respectively, the SM (third-family) left-handed quark doublet and right-handed quark singlets, $\phi$ is the Higgs doublet, and the components in the three field multiplets must be combined in such a way that the term is an $\SU{2}_L$ singlet. When hypercharge conservation is also imposed, only one of these Yukawa terms is allowed for each vector-like multiplet (with a coupling $\lambda$ that can be taken real), except for the doublet $\mathbf{2}_{\frac{1}{6}}$, which can couple to both $t_R$ and $b_R$ (with two real $\lambda$'s and a relative phase). 
In extensions with more than one type of multiplet, it is also possible to write Yukawa terms coupling heavy doublets with heavy singlets or triplets.

\begin{table}[th]
\begin{center}
\caption{Leading corrections to the couplings induced by each vector-like quark multiplet. Up, down and up-down arrows indicate positive, negative or indefinite corrections, respectively, whereas "---" means no corrections. \label{t:1}}
\begin{tabular}{ c  c  c  c  c  c  c  c }
\br
\mbox{} & $T$ & $B$ & $\left( \begin{array}{c} T \\ B \end{array} \right)$ &
$\left( \begin{array}{c} X \\ T \end{array} \right)$  & 
$\left( \begin{array}{c} B \\ Y \end{array} \right)$ & 
$\left( \begin{array}{c} X \\ T \\ B \end{array} \right)$ & 
$\left( \begin{array}{c} T \\ B \\ Y \end{array} \right)$ \\
\mr
\vphantom{\large $\frac{1}{2}$} $V_L$ & $\downarrow$ & $\downarrow$ & --- & --- & --- & $\uparrow$ & $\uparrow$ \\ \hline
\vphantom{\large $\frac{1}{2}$} $V_R$ & --- & --- & $\updownarrow$ & --- & --- & --- & --- \\ 
\hline
\vphantom{\large $\frac{1}{2}$} $X_L^t$ & $\downarrow$ & --- & --- & ---& ---& $\downarrow$ & $\uparrow$ \\ 
\hline
\vphantom{\large $\frac{1}{2}$} $X_R^t$ & --- & --- & $\uparrow$ & $\uparrow$ & --- & --- & --- \\ 
\hline
\vphantom{\large $\frac{1}{2}$} $X_L^b$ & --- & $\downarrow$ & --- & --- & --- & $\uparrow$ & $\downarrow$ \\
\hline
\vphantom{\large $\frac{1}{2}$} $X_R^b$ & --- & ---& $\uparrow$ & ---& $\uparrow$ & --- & --- \\
\hline
\vphantom{\large $\frac{1}{2}$} $Y_V^t$ & $\downarrow$ & --- & $\downarrow$ & $\downarrow$ & --- & $\downarrow$ & $\downarrow$ \\
\br
\end{tabular}
\end{center}
\end{table}
Upon electroweak breaking, the interactions in Eq.~\refeq{Yukawa} give rise to mass terms that mix the heavy and SM quarks. When this matrix is diagonalized, the trilinear couplings of the $t$ and $b$ quarks are modified, giving rise to some of the anomalous couplings in Eqs.~(\ref{anomalous1st}--\ref{anomalouslast}). Equivalently, these corrections can be obtained integrating out the new fields to find $\mathcal{L}^{\mathrm{eff}}$, with scale $\Lambda=M$. In table~\ref{t:1} we show which anomalous couplings are generated by each individual vector-like multiplet at the tree level, and indicate the sign of the corrections. Note that the singlets decrease the value of $V_L$ while the triplets increase it. We also see that three of the multiplets correct some top couplings but do not modify $X_L^b$. This is enforced, without any fine tuning, by gauge invariance.
The anomalous couplings $Y_A$ and $d$ appear at the loop level and have extra $1/(16\pi^2)$ suppressions. There are also exact relations between the different anomalous couplings, which are given in Ref.~\cite{dAPVS2}.

The very same parameters that modify the top couplings enter in the couplings of the heavy quarks to the gauge and Higgs bosons. This allows to connect the searches of new heavy quarks to the precision study of top couplings. On the other hand, the couplings of each extra quark to the $W$, $Z$ and Higgs are also related. These relations imply that their decay branching ratios into different final states depend only on the type of multiplet they belong to, except for the case of the doublet $\mathbf{2}_{1/6}$, in which the branching ratios depend on one continuous parameter~\cite{JAtoppartners}. 
\begin{figure}[th]
\begin{center}
\hspace{0cm} \includegraphics[width=17pc]{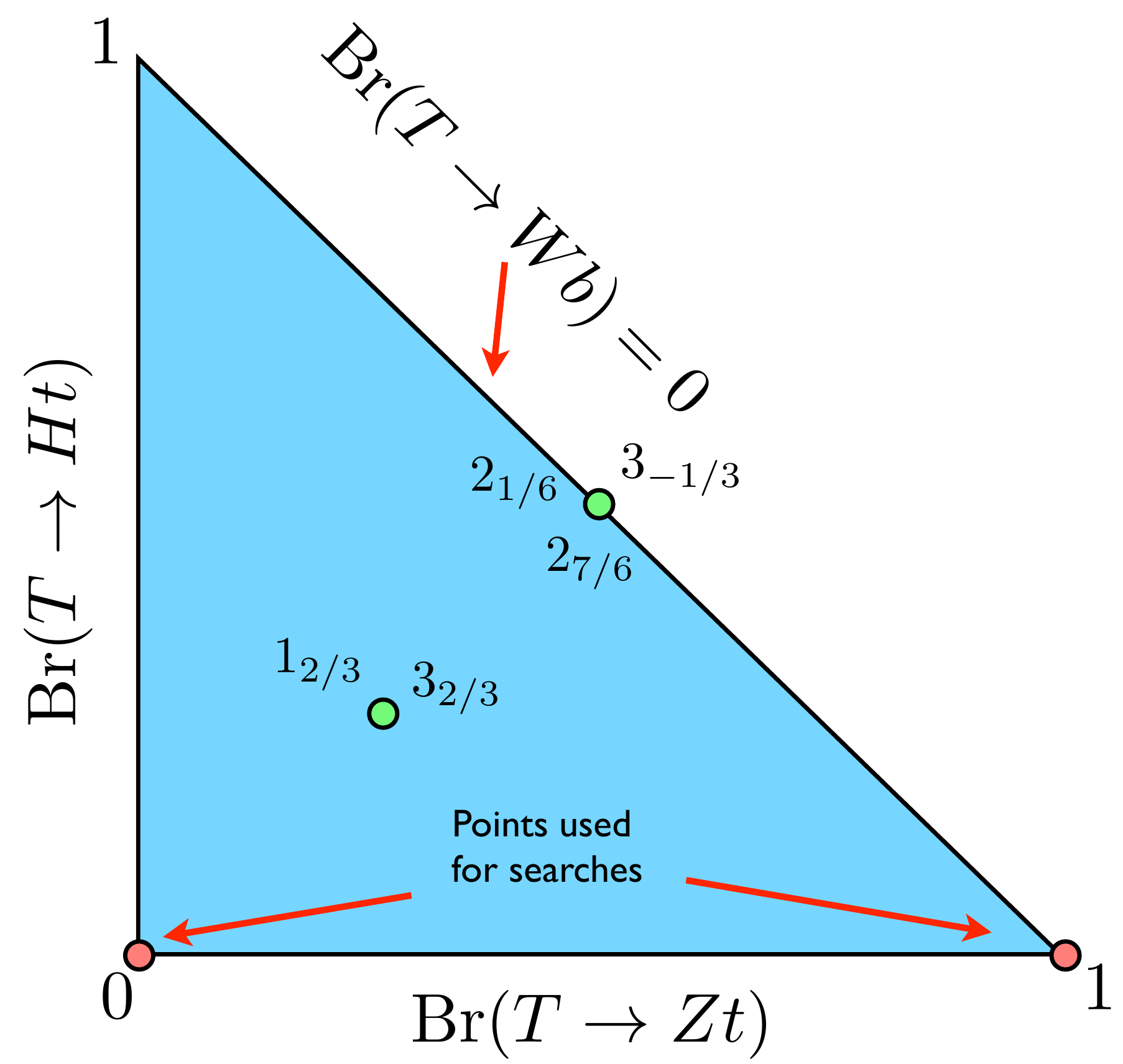} \hspace{0.2cm}
\includegraphics[width=17pc]{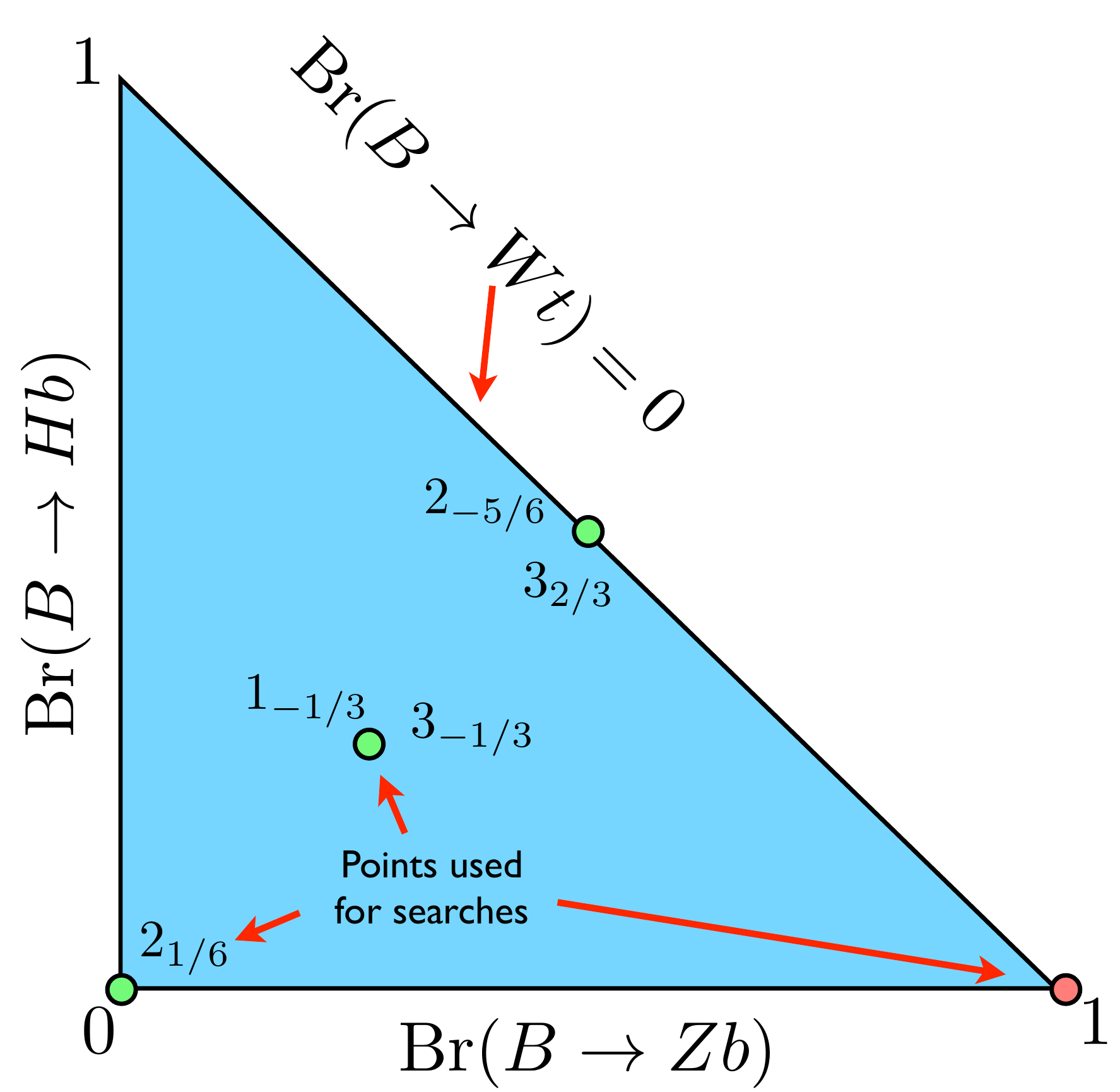}
\end{center}
\caption{\label{f:1} Branching ratios of $T\to Ht$ and $T\to Zt$ (left) and $B\to Ht$ and $B\to Zt$ (right). The green points indicate the branching ratios in the limit of large $M$ for each multiplet, whereas the arrows point to the branching ratios usually assumed in experimental searches.}
\end{figure}
This is illustrated in Fig.~\ref{f:1}, where we represent the possible branching ratios of heavy $T$ and $B$ quarks decaying into final states with $Z$ and Higgs bosons. The branching ratio into $W$ bosons is determined from these by the requirement that $\mathrm{Br}(T\to Zt)+\mathrm{Br}(T\to Ht)+ \mathrm{Br}(T\to Wb)=1$, and analogously for $B$. The green points indicate the branching ratios in the limit of large $M$ for each multiplet. For $\mathbf{2}_{1/6}$, we have assumed that the mixing with the top is much larger than the mixing with the b quark (in general this multiplet would be represented by a line). The arrows point to the branching ratios usually assumed in experimental searches. We see that in many cases these assumptions do not correspond to the allowed branching ratios, so the results of these searches need to be reinterpreted. \footnote{The Atlas collaboration has recently interpreted their searches as constraints on the allowed points of these triangles~\cite{AtlasTriangle}.}

Finally, let us comment on the contribution of these top partners to Higgs production by gluon fussion and Higgs decay into $\gamma \gamma$. The SM predictions are corrected in two ways: by loops of the heavy quarks and by modifications of the top (and $b$ quark) couplings. It is clear that both types of corrections vanish in the limit $M\to \infty$, since the mass of these vector-like fermions does not arise from a Yukawa coupling to the Higgs doublet. Furthermore, it turns out that the corrections from $T$ loops and $t$ anomalous couplings cancel out against each other to a good approximation, for all types of multiplets, as long as there are no Yukawa couplings between the heavy quarks. On the other hand, the loops with $B$ and $b$ quarks are suppressed by $m_b$. To obtain larger corrections, it is necessary to include more than one type of multiplet, as in the models in Ref.~\cite{Bonne:2012im}. Note also that in the effective Lagrangian formalism the effect of the heavy quark loops is reproduced by additional local operators that contribute to these amplitudes. 

\ack
We thank the organizers of Top 2012 for a very pleasant and interesting conference. This work has been supported by the MICINN project FPA2010-17915.

\end{document}